\documentclass[prl,twocolumn]{revtex4-1}
\usepackage{fullpage}
\usepackage{draftcopy}
\usepackage{graphicx}
\usepackage{verbatim}
\usepackage{color}
\usepackage{amsmath}

\newcommand{\swapFn}{\text{S}}
\newcommand{\swapOp}{\widehat{\swapFn}}
\newcommand{\swapFo}{\text{Swap}}

\begin{document}

\title{ Direct calculation of the Entanglement Spectrum in Quantum Monte Carlo with application to \textit{ab initio} Hamiltonians }
\author{Norm M. Tubman}
\author{D. ChangMo Yang}
\affiliation{Department of Physics, University of Illinois, Urbana, Illinois, USA }

\date{\today}
\begin{abstract}
Several algorithms have been proposed to calculate the spatial entanglement spectrum from high order Renyi entropies.  In this work we present an alternative approach for computing the 
entanglement spectrum with quantum Monte Carlo for both continuum and lattice
Hamiltonians.  This method provides direct access to the matrix elements of the spatially reduced density matrix and we determine an estimator that can be used in variational Monte Carlo as well as other Monte Carlo methods.  The algorithm is based on using a generalization of the $\swapFo$ operator, which can be extended to calculate a general class of density matrices that can include  combinations of spin, space, particle and even momentum coordinates.  
   We demonstrate the method by applying it to the Hydrogen and Nitrogen molecules and describe for the first time how the spatial entanglement spectrum encodes a covalent bond that includes all the many body correlations.  
\end{abstract}
\maketitle

\newpage
 Density matrices traced out in real space are becoming a fundamental tool in characterizing different states of matter in condensed matter systems~\cite{op-2,Amico:rmp08,Vidal:prl03,Pollmann:prb10,Qi:prl12,Kitaev:prl06,Flammia:prl09}.   While the interest in spatial reduced density matrices (RDM) is quite recent, the use of density matrices in general is quite ubiquitous~\cite{szabo:book}.  The calculation and usage of particle RDMs in quantum Monte Carlo (QMC) simulations are quite extensive and many techniques have been developed to calculate such density matrices~\cite{holzmann-1,holzmann-2,reptation1}.
 
    Recent QMC entanglement studies have focused on determining the Renyi entropies\cite{renyi-3,renyi-4}.  Calculations of spatial Renyi entanglement with QMC include lattice calculations of topological systems~\cite{renyi-2} and continuum calculations of Fermi liquids~\cite{tubman2,tubman3} and molecules~\cite{tubman1}.  The Renyi  entropies calculated in these works are generally determined with the $\swapFo$ operator which can be applied to interacting systems.  In the community of \textit{ab initio} research,  there has also been much recent work on using QMC to make highly accurate calculations of the momentum distribution of realistic materials~\cite{holzmann-2}.  It turns out the estimators used to calculate the momentum distribution can be seen as a form of the $\swapFo$ operator.  In this work we use the best techniques that have been developed from both of these communities to introduce a generalization of the $\swapFo$ operator, making it an efficient tool to calculate the entanglement spectrum of spatial RDMs. 


   

The spatial entanglement spectrum is derived from a density matrix $\rho_{A}$ in which a system is split into two regions (A and B), and the degrees of freedom in region B are traced out.  The matrix elements of such a density matrix can be expanded in any basis that is complete in region A.  However, in our numerical calculations we in general can only consider a finite number of basis elements.  Therefore practical calculations of the entanglement spectrum are basis dependent, and a carefully selected basis is required.  
 The  algorithm presented here is different from recent proposals~\cite{lau1,assad1,grover1,sling1} for calculating the entanglement spectrum in several ways.  First, there is no calculation of the high order Renyi entropies, and no need to use Maximum Entropy techniques to project out the spectrum.  Additionally, because $\rho_{A}$ is expanded in a basis set as part of our approach, it is possible to select a good basis set to reduce the size of matrix that needs to be constructed.
  
     \textit{Generalized Swap Operator}: Our approach for calculating the entanglement spectrum is to expand the usage of the $\swapFo$ operator, which was first used in QMC for calculating the Renyi entropy of spatial RDMs on a spin lattice~\cite{renyi-1}.  It is based on the replica trick in which coordinates are swapped between copies of a trial wave function.  The $\swapFo$ operator was originally defined in a Hilbert space has been enlarged as a tensor product with itself (although smaller enlargements can be and are used in this work), and we consider its effect when applied to a trial wave function written in the form $\Psi_\text{T} = \sum_{\alpha\beta} C_{\alpha_{}\beta_{}}|\alpha_{}\rangle |\beta_{} \rangle$, where $\alpha$ and $\beta$ are orthonormal basis elements in regions A and B respectively.  With this form of the wave function the $\swapFo$ operator is defined as
\begin{align}
  \swapOp_{A} & \left( \sum_{\alpha_{1}\beta_{1}} C_{\alpha_{1}\beta_{1}}|\alpha_{1}\rangle |\beta_{1} \rangle \right) \otimes \left( \sum_{\alpha_{2}\beta_{2}}D_{\alpha_{2}\beta_{2}} |\alpha_{2}\rangle |\beta_{2} \rangle \right) \notag \\
  {} = & \sum_{\alpha_{1}\beta_{1}} C_{\alpha_{1}\beta_{1}} \sum_{\alpha_{2}\beta_{2}}D_{\alpha_{2}\beta_{2}} |\alpha_{2}\rangle |\beta_{1} \rangle \otimes  |\alpha_{1}\rangle |\beta_{2} \rangle \; .
  \label{eqn:swap1}
\end{align}
Taking the expectation value of the $\swapFo$ operator gives
\begin{align}
  & \left\langle \Psi_{\text{T}} \otimes \Psi_{\text{T}} \left| \swapOp_{A} \right| \Psi_{\text{T}} \otimes \Psi_{\text{T}} \right\rangle \notag \\
  {} & \qquad = \sum_{\alpha_{1}\beta_{1}\alpha_{2}\beta_{2}}C_{\alpha_{1}\beta_{1}}C_{\alpha_{2}\beta_{1}}^* C_{\alpha_{2}\beta_{2}} C_{\alpha_{1}\beta_{2}}^* \notag \\
  {} & \qquad = \sum_{\alpha_{1}\alpha_{2}}(\rho_{A})_{\alpha_{1}\alpha_{2}}(\rho_{A})_{\alpha_{2}\alpha_{1}} = \text{Tr}(\rho_{A}^{2}) \; .
  \label{eqn:swapexp}
\end{align}
The degrees of freedom over which a wave function can be partitioned is not limited to spatial degrees of freedom, and are in fact quite general as there has been some recognition that particle and spatial RDMs can be calculated in similar ways~\cite{sling1,herdman1}.  More generally, in a QMC calculation one can imagine swapping coordinates of spin, space, particle and momentum~\cite{mom1,mom2} in some cases.   Combinations of such degrees of freedom correspond to a ``hybrid reduced density matrix'' are easily accessible, although we are unaware of this being exploited as of yet.  Thus the techniques developed here are not limited to the spatial RDM.   

The term \emph{Swap operator} has not been traditionally used by the QMC community for particle RDM calculation, but the evaluation of such quantities can be thought of as its generalization.  
Of particular interest to this work is a form most recently used in the accurate calculations of the momentum distribution~\cite{holzmann-1},
\begin{align}
  \label{eqn:mom} \rho_{1}(k) & = \left\langle \Psi_\text{T}(\mathbf{R'}) \otimes e^{i\mathbf{k}\cdot\mathbf{r'}} \left| \swapOp \right| e^{-i\mathbf{k}\cdot\mathbf{r}} \otimes \Psi_\text{T}(\mathbf{R}) \right\rangle . 
\end{align}

The evaluation of this expectation value is calculated as an exchange of one particle of the many body trial wave function, $\Psi_\text{T}$, with an electron sampled from a single particle plane wave basis element.  What is important to note here is that the $\swapFo$ operator is being used to project the 1-RDM in a basis set of interest, the plane wave basis.   


By comparing Equations~(\ref{eqn:swapexp}) and (\ref{eqn:mom}), one might expect that we can use the $\swapFo$ formalism to project the spatial entanglement matrix into a basis separately for both region A and region B.  We can see that this can be done explicitly by considering $\swapOp$ acting on a single basis element $|\alpha_{1}\rangle$ in region A,
\begin{align}
  \swapOp_{A} & \left[ |\alpha_{1}\rangle \otimes \left( \sum_{\alpha_{2}\beta_{}} C_{\alpha_{2}\beta_{}}|\alpha_{2}\rangle |\beta_{} \rangle \right) \right] \notag \\
  {} & = \sum_{\alpha_{2}\beta_{}}C_{\alpha_{2}\beta_{}}  |\alpha_{2}\rangle \otimes | \alpha_{1}\rangle |\beta_{} \rangle \; .
\end{align}

We can then evaluate the expectation of the $\swapFo$ operator to calculate the following matrix element,
\begin{align}
  & \left\langle \Psi_{\text{T}} \otimes \alpha_{2} \left| \swapOp_{A} \right| \alpha_{1} \otimes \Psi_{\text{T}} \right\rangle \notag \\
  {} & \qquad = \sum_{\beta_{}}C_{\alpha_{2}\beta_{}}C_{\alpha_{1}\beta_{}}^* = (\rho_{A})_{\alpha_{1}\alpha_{2}} \; .
  \label{eqn:final}
\end{align}

These are the matrix elements for the spatial RDM of which the eigenvalues make up the entanglement spectrum.  The $\alpha$ basis elements are different from the plane waves in Equation~(\ref{eqn:mom}) in that they can involve multiple particles, and they only have support in region A.  
 This equation was derived with a basis set for region A such that $\langle \alpha_{1}|\alpha_{2} \rangle = \delta_{\alpha_{1},\alpha_{2}}$.  
  The estimator in QMC for these matrix elements can be derived as
\begin{widetext}
\begin{align}
 & \left\langle \Psi_\text{T} \otimes \alpha_{i} \left| \swapOp_{A} \right| \alpha_{j} \otimes \Psi_\text{T} \right\rangle  = \int d\mathbf{x}_{1} \cdots d\mathbf{x}_{N} d\mathbf{x}_{N+1} \cdots d\mathbf{x}_{\alpha(N)} \Psi^{*}_\text{T}(x_{A_{1}},x_{B_{}})\alpha^{*}_{i}(x_{A_{2}})\alpha_{j}(x_{A_{1}})\Psi_\text{T}(x_{A_{2}},x_{B_{}}) \nonumber \\
 & \qquad \qquad = \int d\mathbf{x}_{1} \cdots d\mathbf{x}_{N} d\mathbf{x}_{N+1} \cdots d\mathbf{x}_{\alpha(N)} |\Psi_\text{T}(x_{A_{1}},x_{B_{}})|^{2}|\alpha_{i}(x_{A_{2}})|^{2}\frac{\Psi_\text{T}(x_{A_{2}},x_{B_{}})\alpha_{j}(x_{A_{1}})}
{\Psi_\text{T}(x_{A_{1}},x_{B_{}})\alpha_{i}(x_{A_{2}})} \label{eqn:est} ,
\end{align}
\end{widetext}
where $x_{A_{1}}$, $x_{B}$ are the coordinates of electrons in regions A and B sampled from $|\Psi_\text{T}|^{2}$ and $x_{A_{2}}$ are coordinates sampled from $|\alpha^{}_{i}|^{2}$.  Thus, if one had a complete basis set in region A, Equation~(\ref{eqn:final})  is all that is needed in principle to calculate the full entanglement spectrum.  For practical calculations it is expensive to use large basis sets and thus finding a rapidly convergent basis set is important. 

In our implementation we calculate all the matrix elements with equation (\ref{eqn:est}) in a single variational Monte Carlo (VMC) calculation in which the wave function $\Psi_{\text{T}}$ and all the $\alpha_{i}$ are all sampled simultaneously.  At each step a walker position is sampled from $|\Psi_{\text{T}}|^{2}$, exactly as in standard VMC.   We then identify all the $\alpha_{i}$ that are compatible with this walker, in that they must have exactly the same number of spin-up and spin-down electrons in region A.   For all compatible $\alpha_{i}$, where each $\alpha_{i}$ has its own walker, we perform a VMC step that samples from $|\alpha_{i}|^{2}$.   We use the wave function evaluations to calculate the denominator of the estimator in equation (\ref{eqn:est}) and then we swap the region A coordinates  ($x_{A}$) between the walkers of $\alpha_{i}$ and $\Psi_{\text{T}}$ to calculate the numerator of the estimator.

\begin{figure}[tbp]
\centering
\includegraphics[scale=0.1]{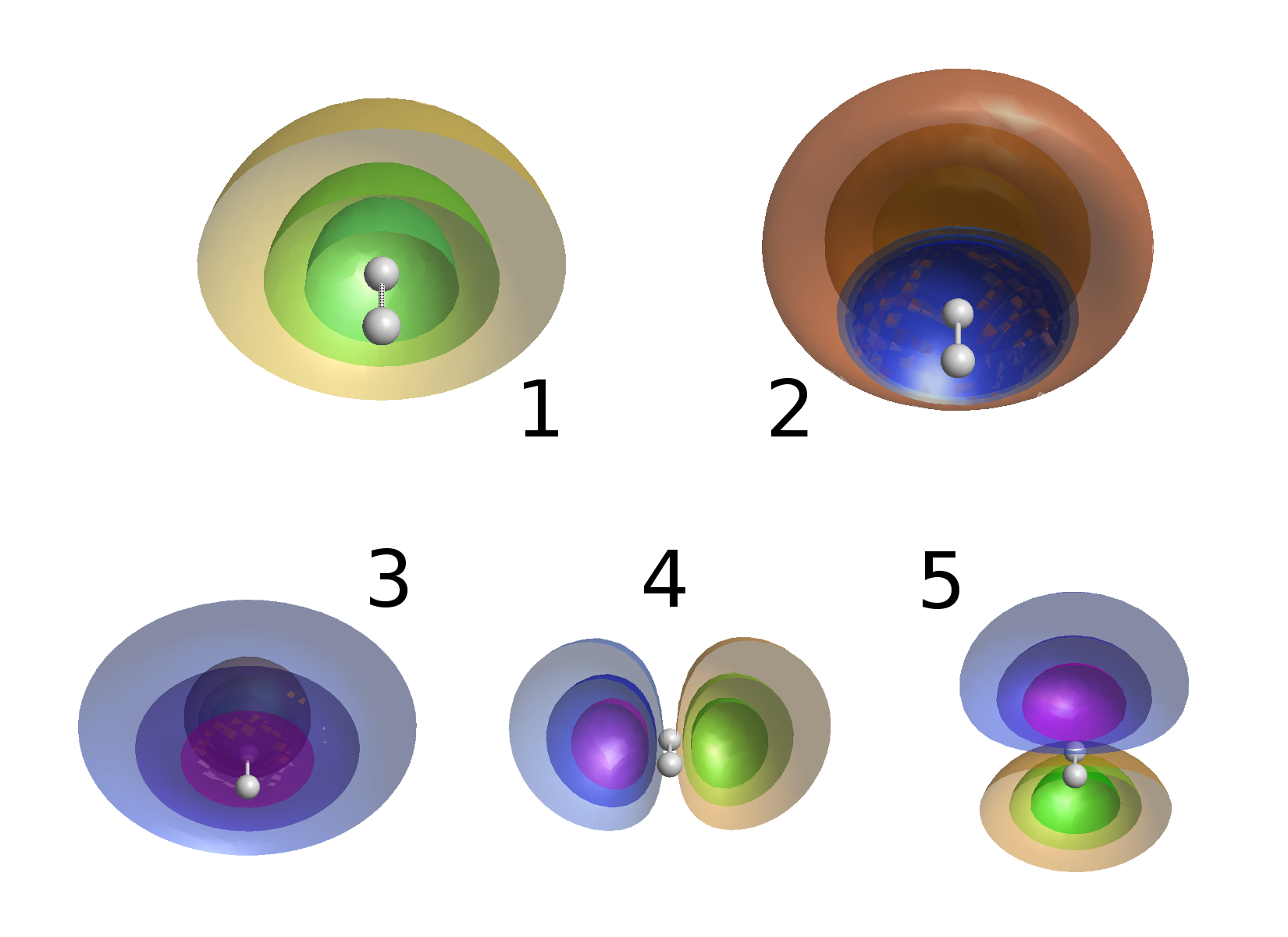}
\caption{The five orbitals with the largest eigenvalues of the effective entanglement Hamiltonian for H$_{2}$.  The orbitals, unlike normal single particle orbitals, exist only in the half space of region A. \label{fig:orbs}}  
\end{figure}

\textit{Efficient basis set generation}: Creating a rapidly converging basis in which to expand the spatial entanglement spectrum is similar to the problem of creating multi-determinant wave functions for continuum Hamiltonians~\cite{szabo:book}, where there are many possible basis states and one has to select which determinants to include in the diagonalization of the Hamiltonian.   A common basis that is used for a multi-determinant expansion is the natural orbital basis.  The natural orbitals are the eigenvectors for the 1-RDM of a many body trial wave function.   Whether or not this is the most compact basis for a multi-determinant expansion is something that has been discussed extensively~\cite{szabo:book} but it is not something that can be proved rigorously as there are notable exceptions~\cite{natorb}. Regardless, it is used in many techniques to diagonalize continuum Hamiltonians~\cite{gamess-1}.



In a manner similar to generating natural orbitals, we suggest a good basis set for such an expansion of the spatial RDM can be generated with the correlation method~\cite{corr-2,redmat-1}.   The correlation method was developed originally to calculate the spatial entanglement for single determinant wave functions.  However, an effective entanglement Hamiltonian from the correlation method can be generated and diagonalized for a multi-determinant wave function.
These eigenvectors can be considered the spatial natural orbitals. This is a natural definition to adopt as the correlation matrix which is used to determine the effective entanglement Hamiltonian is given by
  ${C}_{\alpha_{1}\alpha_{2}} = \text{Tr}({\rho}_\text{1}c^{\dag}_{\alpha_{1}}c_{\alpha_{2}}) $,
where $\alpha$ represents degrees of freedom that exist only in region A.  In other words we are using matrix elements, from the 1-RDM, that exist only in region A to generate our spatial natural orbitals.  The effective entanglement Hamiltonian can be constructed from the correlation matrix as follows, 
\begin{equation}
\mathbf{H}^{(1)}_\text{ent}= \text{ln} \left( \frac{\mathbf{I}-\mathbf{C}}{\mathbf{C}} \right) \; .
\end{equation}


The rank of the effective entanglement Hamiltonian is arbitrarily large since it was derived from $\rho_{1}$ of an interacting system.   We will generally have to limit ourselves to a subset of the eigenvectors of the entanglement Hamiltonian to create our basis.  We pick this subset based on the eigenvalues of the entanglement Hamiltonian.  The eigenvalues, even for a multi-determinant wave function, range between 0 and 1.  For selecting orbitals in our truncated expansion we observe that orbitals with eigenvalues close to 1 are the most important to retain, which is what we expect as this is the rule of thumb for selecting natural orbitals for multi-determinant expansions.


The construction of the $|\alpha\rangle$ basis elements is straightforward once the entanglement Hamiltonian is diagonalized~\cite{peschel-func} and a subset of the eigenvectors is selected.   The $|\alpha\rangle$  are single determinant states that differ from familiar determinant basis sets in that all particle sectors are present.  All the single determinant states of $0$ to $N$ particles should be constructed that are consistent with the physical number of particles in each spin species.

\textit{Spectrum of a covalent bond}:  Ultimately we want to apply these techniques to condensed matter systems but in this work we instead first look at something interesting that has not yet been studied with the entanglement spectrum,  molecular bonding, and in particular, the H$_{2}$ molecule.  The H$_{2}$ molecule is the prototypical covalent bond, and we are interested if the entanglement spectrum can be used to characterize the properties of bonding in molecules~\cite{tubman1}. 
For this system, we take the half space as our spatial partition, dividing the space equally between the two hydrogen atoms.  We note that it is possible to use the algorithms here with any spatial partition of interest. 

\begin{figure}[tbp]
\centering
\includegraphics[scale=0.4]{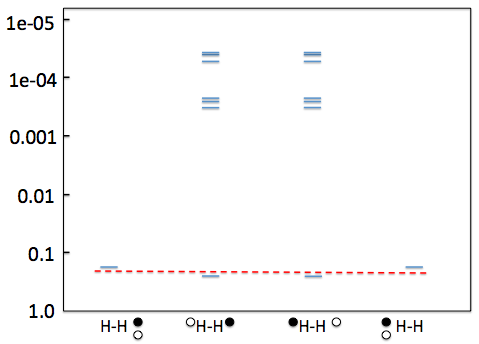}
\caption{The entanglement spectrum of full configuration interaction (blue) ground state wave function of H$_{2}$.  The red dashed line represents the location of he four Hartree-Fock entanglement eigenvalues (0.25).  The four sectors (columns), represent the number of particles and spins in region A.  From left to right the sectors are , ``zero", ``spin-up", ``spin-down", and ``two electrons".   \label{fig:h2-spec}} 
\end{figure}

For a single determinant the spatial entanglement has at most $N$ degrees of freedom which determine how electrons fluctuation through the spatial partition of interest.  However,  when interactions are introduced, the entanglement spectrum can take a more complex form.  In Figure~\ref{fig:h2-spec}, we show the fully interacting entanglement spectrum for $\text{H}_{2}$ and compare to the Hartree-Fock result.  The multi-determinant wave function is a full configuration interaction (CI) calculation with a correlation-consistent basis of penta-zeta  quality (cc-pV5Z)~\cite{c2-2,gamess-1}.  

For a Hartree-Fock wave function, there are four eigenvalues of the entanglement spectrum that is equally split with the value of 0.25, where each state can be labeled by the number of electrons and spins in region A.  These entanglement eigenvalues can be directly interpreted as a probability,  and we can then say there is a 25\% chance for each of the following states in region A: zero electrons, one spin-up electron, one spin-down electron and two electrons.  

For the full CI wave function, the electrons correlations are such that they avoid each other, which can be easily seen in Figure~\ref{fig:h2-spec}.  In particular we can identify how the four states from the non-interacting system evolved into the many body spectrum.  The spectral weight for the 0- and 2-particle sectors are reduced in the interacting entanglement spectrum from 0.25 to 0.20.  This  reflects that the electrons should stay away from each other to reduce the coulomb energy of the system.  On the other hand, the 1-particle states have had their probabilities increased to 0.28.  Additionally there is some extra spectral weight, separated by a gap in the spectrum, that represent other modes of the one electron sectors.   We emphasize that although the spectral weight for the higher energy states in H$_{2}$ is small, these states are required to have statistical correlations between the electrons in regions A and B. 
 We can extend this analysis  by considering the single particle orbitals that are used to construct an eigenvector of interest.  In Figure~\ref{fig:orbs} are the five single particle orbitals with the largest eigenvalues of the entanglement Hamiltonian.  Despite being defined only on the half space, we can identify the first three orbitals as having the symmetry that would go into $\sigma$ bonds, and orbitals 4 and 5 as having the character of $\pi$ bonds.  In the H$_{2}$ entanglement spectrum there are two degenerate eigenvectors in the high energy part above the gap which consist of orbitals 4 and 5.  Thus as far as bonding properties are concerned, the entanglement spectrum yields a description of many body fluctuations and correlations between two regions which can further be organized by the symmetries of the orbitals that create the $\alpha$ basis set. 

What is new here that is not evident in other bonding descriptions is the effect of many body correlations. A nice picture of this can be seen by noting that the eigenvalues/eigenvectors of the entanglement spectrum (for both $\rho_{A}$ and $\rho_{B}$) can be used to generate a Schmidt decomposition of the wave function, $\psi_\text{T} = \sum_{i} c_{i}|\alpha_{i}\rangle |\beta_{i} \rangle$.  This is an exact representation of the wave function in which each basis state represents statistically uncorrelated electrons between regions A and B.  Thus whenever there are single states in a sector separated by a large gap from the rest of the spectrum, as is the case for H$_{2}$, then we can build a picture of bonding from these corresponding low energy states. This is to say we characterize the bonding as electrons fluctuating between regions A and B as in the traditional pictures of bonding.  However, in the case of a small or vanishing gap, many body correlations become important between the electrons in the different regions. 

\begin{figure}[tbp]
\centering
\includegraphics[scale=0.3]{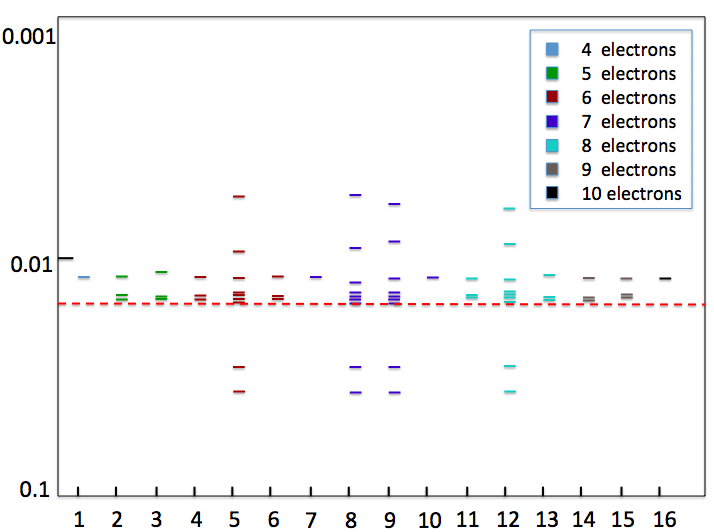}
\caption{The 64 largest values of the entanglement spectrum of a multi-determinant ground state wave function of $\text{N}_{2}$.  The red dashed line represents the location of the 64 largest Hartree-Fock entanglement eigenvalues (0.0156).  The $x$-axis serve as an index for 16 different sectors which are distinguished by particle number and spin polarization in region A.   Sectors 1-8 are symmetric with respect to sectors 9-16 within error bars. \label{fig:N2-spec}} 
\end{figure}

As a demonstration of the method applied to a larger systems, we show in Figure~\ref{fig:N2-spec} the entanglement spectrum of $\text{N}_{2}$, which has 14 electrons.  There is a lot of interesting effects represented in the spectrum, the most important of which is the clear formation of bands in the spectrum.   The Schmidt decomposition requires that pairs of sectors have equivalent eigenvalues, such as (8,~9) and (5,~12) but it is surprising that all 4 of these sectors (5,~8,~9,~12) are equivalent.   Even more surprising is that several eigenvalues show up multiple times in the different sectors, effectively forming bands.  We suggest that these bands can be considered many body resonances that are analogous to delocalized bonding orbitals.  A full description of this spectrum as well as the underlying physics will be described in a future publication.

We mention briefly about the quality of our spatial natural orbital basis set.  For both H$_{2}$ and Li$_{2}$ (another molecule we tested), we find the the first four basis elements generated in our method capture more than 99\% of the spectral weight of $\rho_{A}$, and that the next few basis elements captures a large majority of the remaining weight.   For N$_{2}$ the first 1024 basis elements capture more than 97\% of the spectral weight.  We can say that in the limiting case of Hartree-Fock wave functions is that the spatial natural orbital basis set consists of the exact eigenvectors of $\rho_{A}$ and therefore is the most efficient basis set that can be used.  Thus whenever a system is only weakly interacting we expect the method to work especially well for generating basis sets.  In the case of  Fermi-liquids this means the basis sets are likely to be efficient in the high density limit~\cite{tubman2,tubman3}.  For strongly interacting systems, such as transition metals molecules and low density Fermi-liquids, the spatial natural orbitals are likely to remain a good single determinant basis set for the expansion, but the number of basis elements required to converge the spectrum might be large.  Of course if methods are developed to build better multi-determinant basis sets, they could be used directly in the method presented here.

In this letter we have proposed a method by which the entanglement spectrum can be calculated in QMC.  Furthermore, we have shown that this method can be used efficiently with a spatial natural orbital basis.  We expect that this method will not only be useful for condensed matter systems but also for chemistry and the study of bonding.  We have demonstrated our method on the $\text{H}_{2}$ and $\text{N}_2$ molecules and have shown for the first time what a covalent bond looks like through the entanglement spectrum.  In addition it is clear there is still much to be explored with these methods.  Entanglement partitioning and multi-determinant localized orbitals (from the eigenvectors of the spatial RDM) may be useful as many-body generalizations of Bader analysis\cite{bader:book} and Wannier orbitals respectively.  The techniques described here can be used to study and benchmark density matrix embedding theory techniques in the continuum\cite{dmet1,dmet2}.  Although we apply this method only in VMC here, in principle it can be applied with a mixed estimator in fixed-node diffusion Monte Carlo~\cite{holzmann-1} and release-node QMC~\cite{tubman4,tubman5}, in which fermion solutions can be sampled exactly.  Additionally a pure estimator can be sampled with forward walking\cite{mcbook} and reptation Monte Carlo\cite{reptation1}.  

\textit{Acknowledgments}: We would like to thanks David Ceperley, Jeremy McMinis, and Lucas Wagner for useful discussions.  This work was supported by DARPA-OLE program and DOE DE-NA0001789.  This work used the Extreme Science and Engineering Discovery Environment (XSEDE), which is supported by National Science Foundation Grant No. OCI-1053575.

\newpage

 \bibliography{propref3}{}

\end{document}